\documentclass[journal=aamick,manuscript=article]{achemso}
\usepackage[version=4]{mhchem} 
\usepackage{tabularray}
\usepackage{amsmath}
\usepackage{epsfig}
\usepackage{epstopdf}
\usepackage{multirow}
\usepackage{graphicx}
\usepackage{subcaption}
\usepackage{siunitx}

\author{Nguyen Tran Gia Bao}
\affiliation{Faculty of Physics and Physics Engineering, University of Science, Ho Chi Minh City 700000, Viet Nam}
\altaffiliation{Vietnam National University, Ho Chi Minh City 700000, Viet Nam}
\author{Ton Nu Quynh Trang}
\affiliation{Faculty of Physics and Physics Engineering, University of Science, Ho Chi Minh City 700000, Viet Nam}
\altaffiliation{Vietnam National University, Ho Chi Minh City 700000, Viet Nam}
\author{Nam Thoai}
\affiliation{Ho Chi Minh City University of Technology, Ho Chi Minh City 700000, Viet Nam}
\altaffiliation{Vietnam National University, Ho Chi Minh City 700000, Viet Nam}
\author{Phan Bach Thang}
\affiliation{Center for Innovative Materials and Architectures (INOMAR), Ho Chi Minh City 700000, Viet Nam}
\altaffiliation{Vietnam National University, Ho Chi Minh City 700000, Viet Nam}
\author{Vu Thi Hanh Thu}
\email{vththu@hcmus.edu.vn}
\affiliation{Faculty of Physics and Physics Engineering, University of Science, Ho Chi Minh City 700000, Viet Nam}
\altaffiliation{Vietnam National University, Ho Chi Minh City 700000, Viet Nam}
\author{Nguyen Tuan Hung}
\email{nguyen.tuan.hung.e4@tohoku.ac.jp}
\affiliation{Frontier Research Institute for Interdisciplinary Sciences, Tohoku University, Sendai, 980-8578 Japan}

\title[title]
  {Rational Design Heterobilayers Photocatalysts for Efficient Water Splitting Based on 2D Transition-Metal Dichalcogenide and Their Janus}

\abbreviations{}
\keywords{Janus TMDC, photocatalyst, water-spliting, carrier mobility,  Fr\"{o}hlich interaction}

\begin{document}

\begin{abstract}

Direct Z-scheme heterobilayers with enhanced redox potential are viewed as promising for solar-driven water splitting, arising from the synergy between intrinsic dipoles in Janus materials and interfacial electric fields across the layers. This study explores 20 two-dimensional Janus transition-metal dichalcogenide (TMDC) heterobilayers for efficient water splitting. Using density-functional theory (DFT) calculations, we screen them based on band gaps and intrinsic electric fields to identify promising candidates, then further assess carrier mobility and surface chemistry to fully evaluate their overall performance. 
By examining the alignment of synthetic and internal electric fields, we distinguish between Type-I, Type-II, and Z-scheme configurations, enabling the targeted design of optimal photocatalytic materials. Furthermore, we employ the Fr\"{o}hlich interaction model to quantify the mobility contributions from the longitudinal optical phonon mode, providing detailed insights into how carrier mobility, influenced by phonon scattering, affects photocatalytic performance. Our findings demonstrate the potential of Janus-based Z-scheme systems to overcome existing limitations in photocatalytic water splitting by optimizing the electronic and structural properties of 2D materials, highlighting a viable pathway for advancing clean energy generation through enhanced photocatalytic processes.

\end{abstract}

\begin{tocentry}

\begin{center}
\includegraphics[width=0.8\textwidth]{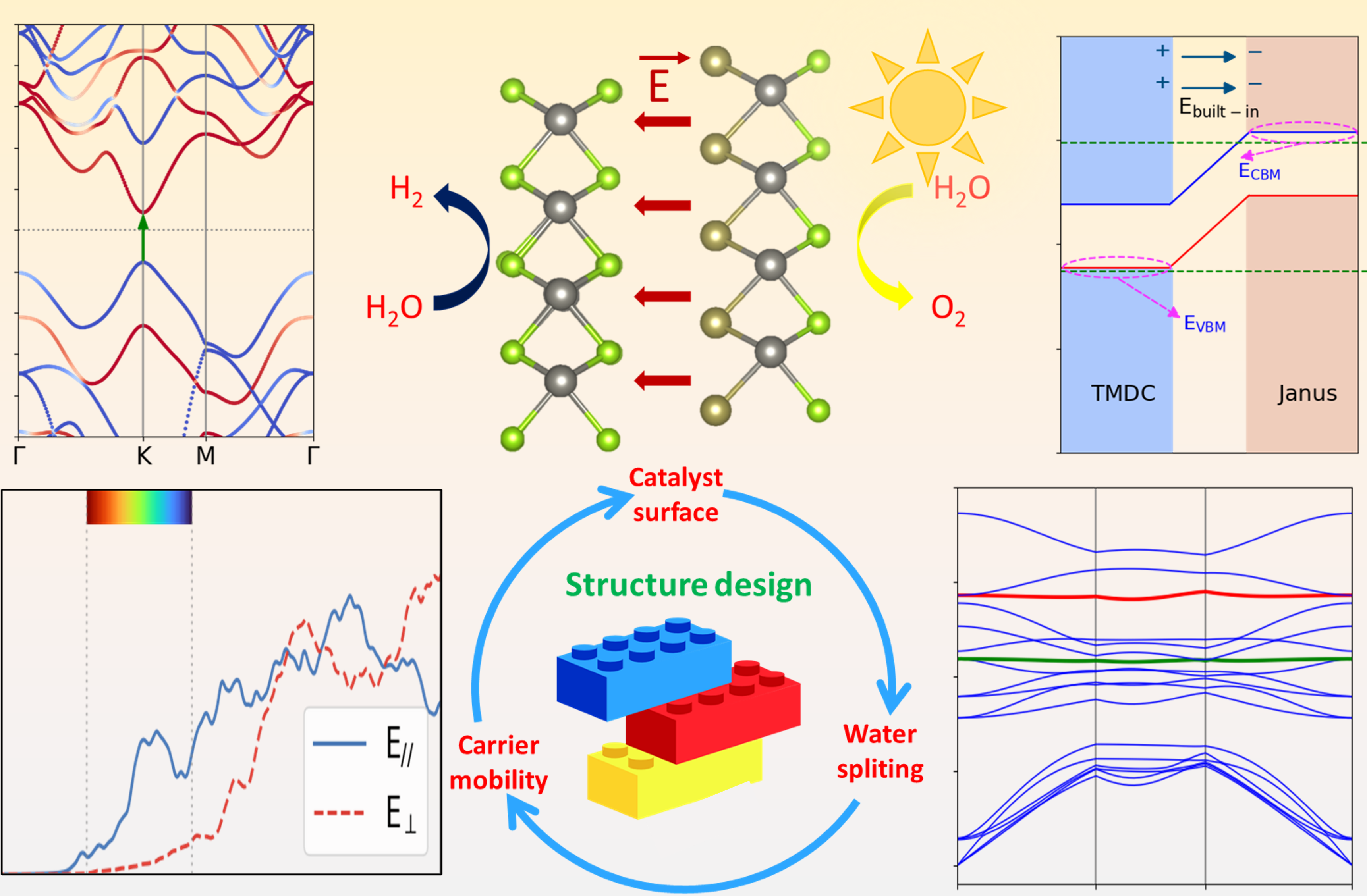}
\end{center}

\end{tocentry}

\section{Introduction}
Photocatalytic water splitting that converts solar energy through the splitting process of water to hydrogen H\(_2\) energy has garnered significant attention in global scientific and industrial communities~\cite{thanh2023janus, trang2024manipulating, trang2024multifunctional}. 
Among the various photocatalysts, 2D materials stand out for water splitting due to their large surface area, quantum confinement, strong light absorption, enhanced charge separation, and structural flexibility \cite{Wang2024-hb, Ma2024-jp}. Their planar structure features a high density of surface atoms and ample space for co-catalyst integration, boosting available active sites and overall efficiency \cite{li2019photocatalyst}. Moreover, the reduced thickness of 2D materials shortens charge-carrier travel distances to the surface, mitigating electron-hole recombination and preserving more carriers for catalytic reactions. These characteristics make 2D materials promising photocatalysts, particularly TMDCs, which benefit from efficient visible light absorption and high charge-carrier mobility for enhanced charge separation \cite{Roy2024-vq}. The TMDCs have demonstrated good chemical stability and robust light-matter interactions, further improving their efficacy in light-driven processes. Despite these strengths, TMDCs are facing plausible limitations in photocatalytic hydrogen generation due to high exciton recombination and strong electrostatic coulomb interactions, leading to inefficient photogenerated charge carriers~\cite{yuan2023first}. 

In comparison with conventional TMDCs, Janus TMDCs include different chalcogen atoms, breaking the out-of-plane symmetry~\cite{hung2024symmetry}. Thus, the Janus TMDCs will create out-of-plane dipoles between the top and bottom layers that could generate an intrinsic electric field and enhance the photocatalytic performance in Janus-TMDCs \cite{Yuan2021-ab, van2020first}. Based on the above reasons, the heterobilayers combining TMDCs and their Janus materials that have been successfully synthesized \cite{hung2023nonlinear, zhang2021spectroscopic} could benefit photocatalytic water splitting. In monolayer 2D material, both the hydrogen evolution reaction (HER) and the oxidation evolution reaction (OER) occur on the same surface of the 2D material, resulting in a higher recombination rate of photogenerated charge carriers \cite{qian2023electronic, liu2023designing}. A promising approach to overcome this limitation is the use of van der Waals (vdW) heterobilayers composed of two semiconductor layers.

These heterobilayers exhibit three primary types of heterostructure alignments: Type-I, Type-II, and Z-scheme, distinguished by their electronic band alignments and charge carrier dynamics. In the Type-I configuration, the conduction band minimum (CBM) and valence band maximum (VBM) of one semiconductor are entirely enclosed within the bandgap of the other \cite{wang2019particulate}. This alignment facilitates the transfer of both electrons and holes to a single material, making it suitable for applications requiring charge recombination, such as light emission.\cite{Bae2021-in} However, it is less effective for photocatalysis due to its limited ability to separate charge carriers. 
In contrast, the Type-II and Z-scheme heterobilayerss offer superior solutions. These configurations distribute the VBM and CBM across both semiconductor layers \cite{Shabbir2024-bc}. 
In particular, the Z-scheme junction features a suitable band alignment and a wide directional built-in electric field. As a result, photogenerated charge carriers with a large redox potential are well-preserved and spatially separated within the Z-scheme heterobilayers. Combining two narrow band gap semiconductors into a Z-scheme configuration could solve the issue of the Type-II configuration, in which the redox ability and light absorption region may not be obtained simultaneously \cite{Liu2023-ld, Gao2022-av, Shabbir2024-bc}. The Z-scheme configuration allows HER and OER activity on separate surfaces \cite{Wang2024-tt}. 

On the other hand, the heterobilayer with the Janus structure generates an intrinsic electric field, which not only enhances the photocatalytic performance but also accelerates the reaction kinetics by forming synergistic surface properties that allow selective adsorption of water molecules on each surface~\cite{cavalcante2019enhancing, xu2022enhanced}. 
Moreover, the configuration of the heterobilayer can determine whether photogenerated carriers follow a Z-scheme or a traditional Type-II pathway \cite{xu2022enhanced,Dong2023-oo}.
However, the diversity of possible Janus-TMDC combinations in heterobilayers poses a significant challenge in determining the optimal configuration for photocatalytic water splitting. 
In this study, we provide a theoretical framework for rapidly identifying effective Janus-TMDC pairs through a top-down approach: candidates are initially filtered based on band alignment, and further refinement follows as needed. Another key challenge is the intrinsic polarization of Janus materials, which uniquely affects their electrical mobility. However, this factor is often neglected in previous Janus or TMDC studies for photocatalytic \cite{cheng2018limits, thanh2023janus,ju2020janusa}. To address this, we have adopted Fr\"{o}hlich interaction model \cite{sohier2016two, cheng2018limits} to account for the mobility adjustments in Janus heterobilayers accurately.

In this paper, we investigate the photocatalytic properties of the 20 configurations Janus-XMY/TMDC-NZ$_2$  heterobilayers, where \( \text{M}, \text{Z} = \text{Mo}, \text{W} \); and \( \text{X}, \text{Y}, \text{Z} \) are chalcogenide elements S, Se, or Te (\( \text{X} \neq \text{Y} \)). 
Using the density functional theory (DFT) calculations, our study focuses on how the interfacial side of the Janus materials (i.e., S, Se, or Te atomic layers) and the different combinations with the TMDCs influence the intrinsic built-in electric field developed across them, affecting their band gap energy, CBM and VBM levels, type of configuration, and particularly the band alignment type suitable for photocatalytic water splitting reactions. Several critical properties closely associated with water-splitting performance, such as STH efficiency, charge transport, and surface chemical reactions, are thoroughly evaluated. In addition to these evaluations, we also introduce a simple way to predict band alignment types based on electronegativity differences and interfacial atomic configurations. 

\section{Theoretical Approach and Computational Details}
In this research, we employed DFT for calculations using the Quantum ESPRESSO package \cite{QE-2017}. The generalized gradient approximation (GGA) with the Perdew-Burke-Ernzerhof (PBE) exchange-correlation functional was utilized as implemented in the standard solid-state pseudopotentials (SSSP) library  \cite{ prandini2018precision, lejaeghere2016reproducibility}. Van der Waals interactions were incorporated using the Grimme DFT-D2 method \cite{thonhauser2015spin, berland2015van}. To address the underestimation by PBE in predicting band edge positions and band gaps, we used the Heyd-Scuseria-Ernzerhof (HSE) hybrid-functional with a mixing parameter 25\%  \cite{heyd2003hybrid, mostofi2008wannier90}. The kinetic cutoff energy for wavefunctions ($E_\text{cut}$) was set to 800 eV, and the Monkhorst-Pack k-point grid was set to $12 \times 12 \times 1$ for PBE and $6 \times 6 \times 1$ for HSE based on the convergence test. All structures were fully relaxed using the BFGS scheme, with convergence criteria for force and energy differences set to $0.25 \times 10^{-3}$ eV/\AA\ and $1 \times 10^{-5}$ eV, respectively. To prevent interlayer interactions, a relatively large vacuum layer of 30 \AA\ was considered along the $z$-direction. To examine the Gibbs free energy and the mobility of the hydrogen atom, we adopted a $3 \times 3 \times 1$ supercell with 54 atoms. For the deformation potential theory (DPT) carrier mobility calculation, a rectangle $\sqrt{3} \times 1 \times 1$ supercell was used to obtain the effective mass of carriers along $x$- and $y$-directions. In the calculation of mobility driven by Fröhlich scattering, the Born effective charge serves as a crucial parameter. It is determined by incorporating long-range electrostatic interactions through density functional perturbation theory \cite{Debernardi1995-jh}. The Wannier90 code and the VESTA software were used for pre-processing and post-processing \cite{pizzi2020wannier90, momma2011vesta}.

\begin{figure}[t]
  \centering
  \includegraphics[width=0.7\linewidth]{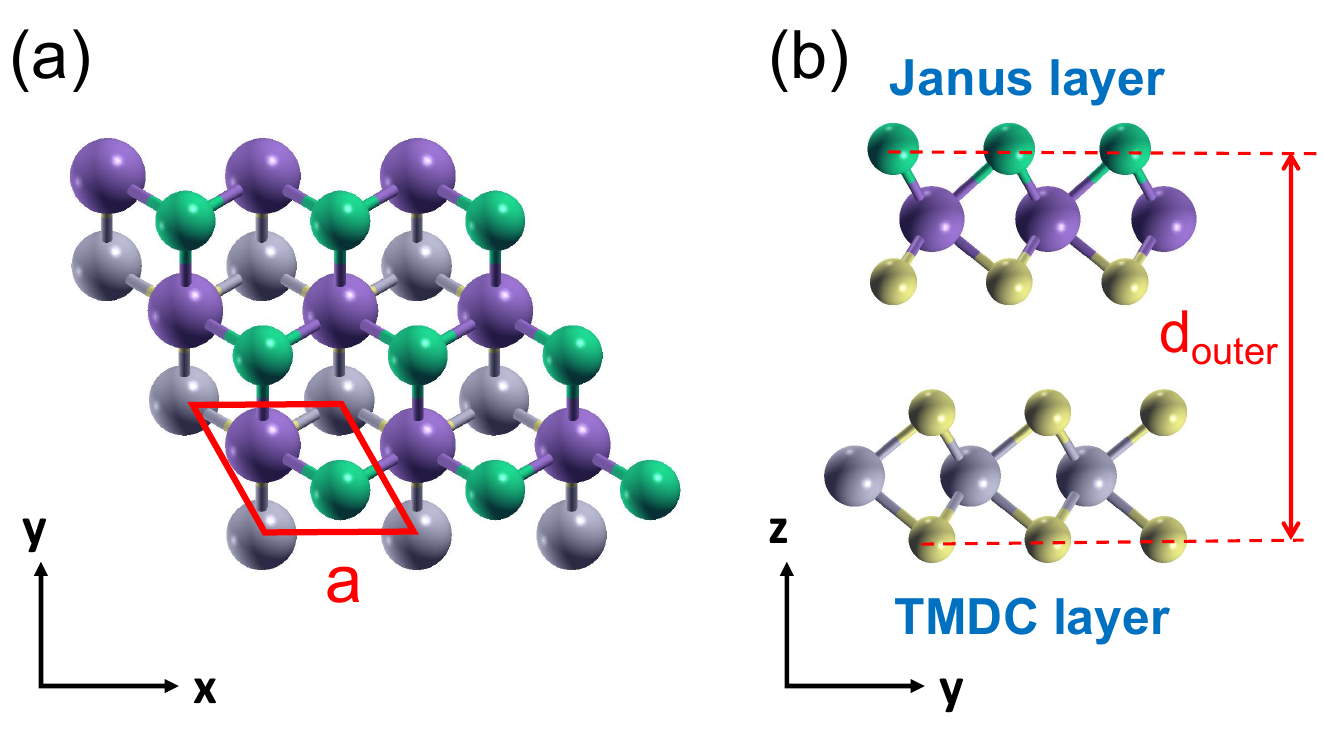}
  \caption{Visual representations of the vdW Janus-TMDC structure: (a) top view and (b) side view.}
  \label{fig:1}
\end{figure}

\section{Results and Discussion}
\subsection{Geometric and Electronic Structures}

The geometric structures of these 20 materials are designed based on the previously synthesized AA stacking order of the MoSSe-MoS$_2$ structure~\cite{zhang2020enhancement}, systematically replacing the metal and chalcogenide atoms, as shown in Figure~\ref{fig:1}. In addition to the AA-stacked mode, the heterobilayers of TMDC/Janus heterostructures also exist in the AB-stacked mode. However, Nguyen et al. \cite{hung2024symmetry, hung2023nonlinear} showed that the linear optical properties and electronic band structures of the AA- and AB-stacking orders are almost identical, which suggests that the photocatalytic characteristics of these orders should be the same. Thus, in the present study, we focus on the AA-stacking for 20 heterobilayers.
The optimized lattice parameters of these heterobilayers are listed in Table~\ref{tab:1}. We note that the structures of MoS$_2$-TeMoSe and WS$_2$-TeWSe did not converge during the optimization calculations in this study. This non-convergence might come from the significant differences in electronegativity between the S and Te atoms at the interface.

\begin{table}[t]
\centering
\caption{Calculated band gap energies $E_g$, work function differences $\Delta \Phi$, alignment type, and effective band gaps $E_\text{eff}$ for various Janus-TMDC heterobilayers.}
\begin{tabular}{lcccc} 
\hline
\textbf{Sample} & \textbf{$E_g$} & \textbf{$\Delta \Phi$} & \textbf{Alignment}& \textbf{$E_\text{eff}$}\\
 & \textbf{(eV)} & \textbf{(eV)} &  \textbf{Type}& \textbf{(eV)}\\
\hline
MoS$_2$-SeMoTe & 0.32 & 0.82 & Type-II & - \\
MoS$_2$-TeMoSe & - & - & - & - \\
MoSe$_2$-SeMoTe & 1.04 & 0.84 & Type-I & 1.04 \\
MoSe$_2$-TeMoSe & 0.48 & -0.60 & Z-scheme & 1.08 \\
MoS$_2$-SMoSe & 1.31 & 0.82 & Type-I & 1.31 \\
MoS$_2$-SeMoS & 0.68 & -0.64 & Z-scheme & 1.32 \\
MoSe$_2$-SMoSe & 0.49 & 0.71 & Z-scheme & 1.20 \\
MoSe$_2$-SeMoS & 1.11 & -0.73 & Type-I & 1.11 \\
WS$_2$-SWSe & 1.44 & 0.79 & Type-I & 1.44 \\
WS$_2$-SeWS & 0.82 & -0.66 & Z-scheme & 1.48 \\
WSe$_2$-SWSe & 0.60 & 0.69 & Z-scheme & 1.29 \\
WSe$_2$-SeWS & 1.21 & -0.70 & Type-I & 1.21 \\
WS$_2$-SeWTe & 0.41 & 0.93 & Type-II & - \\
WS$_2$-TeWSe & - & - & - & - \\
WSe$_2$-SeWTe & 1.10 & 0.80 & Type-I & 1.10 \\
WSe$_2$-TeWSe & 0.97 & -0.63 & Z-scheme & 1.60 \\
MoS$_2$-SWSe & 1.56 & 0.83 & Type-II & 0.73 \\
MoS$_2$-SeWS & 0.95 & -0.58 & Z-scheme & 1.53 \\
WS$_2$-SMoSe & 0.94 & 0.79 & Z-scheme & 1.72 \\
WS$_2$-SeMoS & 1.52 & -0.71 & Z-scheme & 2.23 \\
\hline
\end{tabular}
\label{tab:1}
\end{table}

\begin{figure*}[t]
  \centering
  \includegraphics[width=1\linewidth]{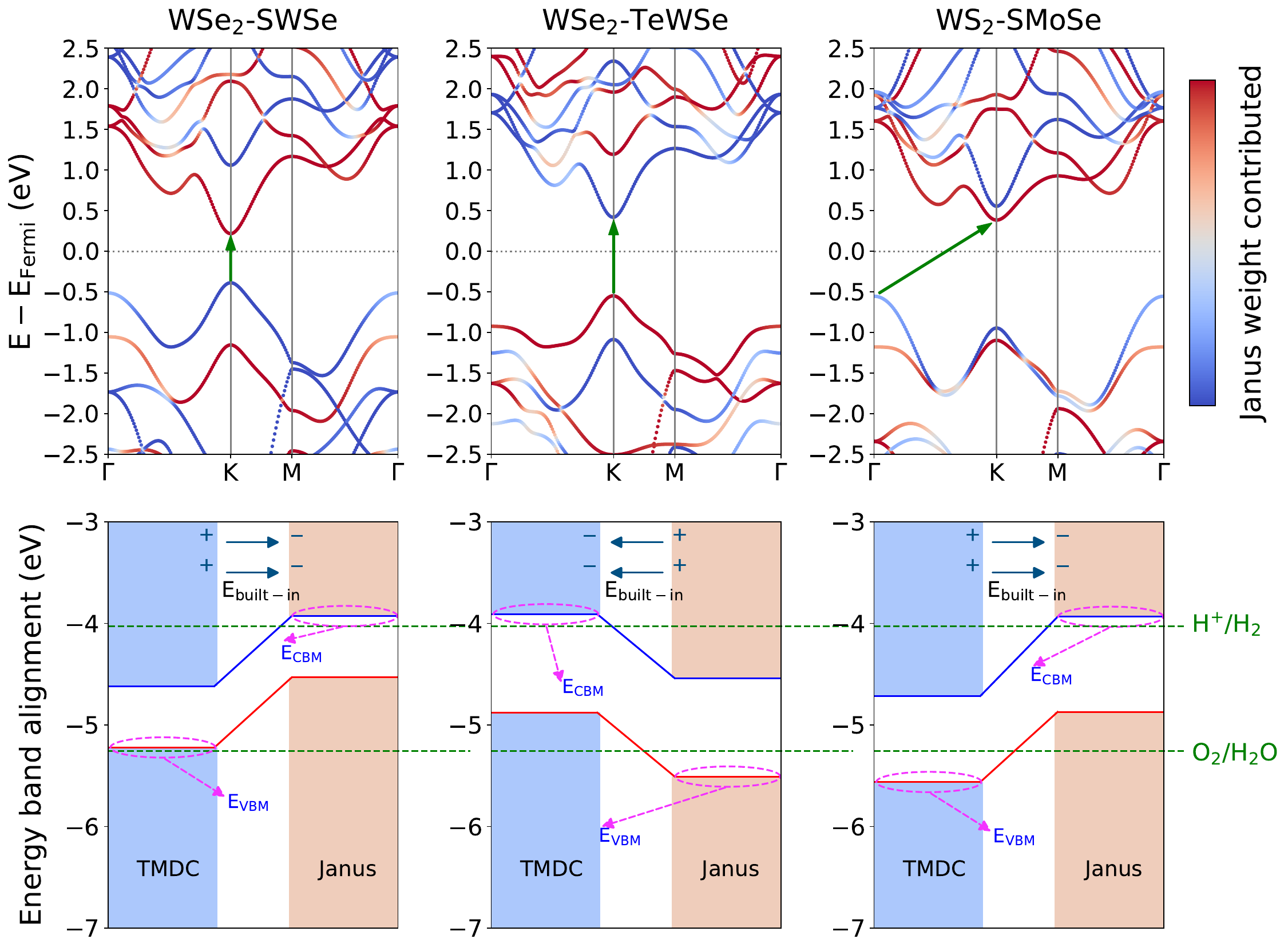}
  \caption{HSE band structure and band alignment at pH = 7 for selected Janus-TMDC heterobilayers: WSe$_2$-SWSe, WSe$_2$-TeWSe, and WS$_2$-SMoSe.}
  \label{fig:2}
\end{figure*}

\subsection{Photocatalytic characteristic}
After completion of structural optimizations, we evaluated the potential of 18 materials as photocatalysts (see Table~S1 in Supporting Information). To determine the electronic properties critical for photocatalytic activity, we performed partial energy band structure for 18 material combinations, in which the partial band structures show the contributions of individual layers in the heterobilayers to the electronic states near the band edges (see Figure~S1, Supporting Information).  By projecting the electronic states onto the atomic orbitals of the Janus and TMDC layers, we identified which layers predominantly contribute to the VBM and CBM. This approach allowed us to classify the materials into Type~I band alignment, where both the VBM and CBM are contributed by the same layer, or Type~II alignment and Z-scheme alignment,  where they are contributed by different layers, based on the spatial localization of their VBM and CBM. 

As shown in Figure~S1 (Supporting Information), 11 materials exhibit a direct band gap, with both the CBM and the VBM located at the $\text{K}$ point of the Brillouin zone. While 7 materials show the indirect band gap with the VBM is found along the $\text{K}$–$\Gamma$ path. The calculated band gap energies for all materials are summarized in Table~S1 (Supporting Information), providing a detailed comparison across the different structures. Here, we employed the HSE hybrid functional for all materials to correct the band gap values, as the commonly used generalized gradient approximation with the GGA-PBE functional tends to underestimate the energy band gap~\cite{hung2022quantum}.

To be considered as a photocatalyst for water splitting, the theoretical band gap should be larger than 1.23 eV to overcome the reduction and oxidation potential of H$_2$O \cite{thanh2023janus, nguyen2022tunable}. However, due to the difference in electrostatic potential between the layers contributing to the VBM and CBM, even materials with smaller band gaps can act as photocatalysts. As shown in Table~S1 (Supporting Information), the materials with Type-I band alignment, including MoSe$_2$-SeMoTe, WSe$_2$-SeWS, and WSe$_2$-SeWTe, have band gaps less than 1.23 eV. Thus, they are unsuitable for water-splitting applications.

For the remaining materials, we calculated the electrostatic potential curves, as shown in Figure~S2 in Supporting Information. The calculations include a dipole correction to suppress artificial dipole fields by introducing an additional ramp-shaped potential~\cite{hung2022quantum}. The intrinsic electric field generated by the Janus layer in the heterobilayers results in a significant potential difference between the two layers, leading to a difference in work function ($\Delta\Phi$) as calculated and presented in Table~\ref{tab:1}. Note that $\Delta\Phi$ is determined by taking the difference between the work functions of the TMDC and Janus layers, following this equation \cite{yuan2023first}: 
\begin{equation}
    \Delta\Phi =  V_{\text{TMDC}} - V_{\text{Janus}},
\end{equation}
where $V_{\text{TMDC}}$ and $V_{\text{Janus}}$ are the vacuum energies of the TMDC and Janus layers in the heterobilayers, respectively. The values of $V_{\text{TMDC}}$ and $V_{\text{Janus}}$ of each heterobilayer are shown in Figure~S2 on the Supporting Information.

The band alignment plays a critical role in determining the photocatalytic properties of Janus-TMDC heterobilayers. When the work function difference between the layers, denoted as $\Delta\Phi$, is positive ($\Delta\Phi > 0$), this indicates that the vacuum level of the TMDC layer is higher than that of the Janus layer. In such a configuration, the TMDC layer predominantly contributes to the VBM, while the Janus layer largely influences the CBM. This arrangement leads to a Z-scheme band alignment, which effectively broadens the separation between the CBM and VBM, enhancing charge carrier separation and improving photocatalytic activity (e.g., WSe$_2$-SWSe and WS$_2$-SMoSe, as shown in Figure~\ref{fig:2}) \cite{Liu2023-ld, Gao2022-av, Shabbir2024-bc}. Conversely, when $\Delta\Phi$ is negative ($\Delta\Phi < 0$), meaning the vacuum level of the Janus layer is higher than that of the TMDC layer, the Janus layer mainly contributes to the VBM, and the TMDC layer to the CBM. This configuration also results in a Z-scheme alignment (e.g., WSe$_2$-TeWSe, as shown in Figure~\ref{fig:2}).

In cases where $\Delta\Phi$ is negligible or does not meet the conditions above, the system exhibits a Type-II band alignment. The effective band gap in a Z-scheme configuration can be expressed as $E_{\text{eff}} = E_g + \Delta\Phi$, where $E_g$ represents the intrinsic band gap, and $\Delta\Phi$ reflects the built-in potential difference at the heterobilayers interface. Materials with $E_{\text{eff}} > 1.23$ eV are considered promising candidates for photocatalytic applications, and their performance is further evaluated through calculations of STH efficiency. The $E_{\text{eff}}$ values for the materials studied are summarized in Table~\ref{tab:1}. We note that the missing $E_\text{eff}$ values of WS$_2$-SeWTe and MoS$_2$-SeMoTe are due to the larger differences in vacuum potential, which result in CBM being positioned below the OER potential and VBM being above the HER potential. This placement makes the electron and hole at CBM and VBM recombination before doing HER and OER processes. As a result, these materials are unsuitable for water splitting.

\subsection{Intrinsic strategies for band alignment control in Janus/TMDC heterobilayers}
Designing heterobilayers of Janus and TMDC materials necessitates the control of interlayer electric fields to tailor band alignments for specific functional applications. Our findings indicate that manipulating these electric fields is key to achieving the desired type of band alignment. To investigate this, we calculate the electric field of the Janus/TMDC heterostructure through the electrostatic potential along the $z$-direction, as shown in Figure~S2 (Supporting Information). The potential difference
between the two chalcogenide layers inside $\Delta V_{\text{inner}}$ and outside $\Delta V_{\text{outer}}$ heterobilayers are defined as follows:
\begin{equation}
    \Delta V_{\text{inner}} = V_{\text{TMDC-inner}} - V_{\text{Janus-inner}},
\end{equation}
and
\begin{equation}
    \Delta V_{\text{outer}} = V_{\text{TMDC-outer}} - V_{\text{Janus-outer}},
\end{equation}
where $V_{\text{TMDC-inner}}$ and $V_{\text{Janus-inner}}$ are potential energies at the top chalcogenide of the TMDC layer and bottom chalcogenide of the Janus layer (see Figure~\ref{fig:1} (b)), respectively. $V_{\text{TMDC-outer}}$ and $V_{\text{Janus-outer}}$ are potential energies at the bottom chalcogenide of the TMDC layer and top chalcogenide of the Janus layer, respectively.

\begin{figure*}[t]
  \centering
  \includegraphics[width=1\linewidth]{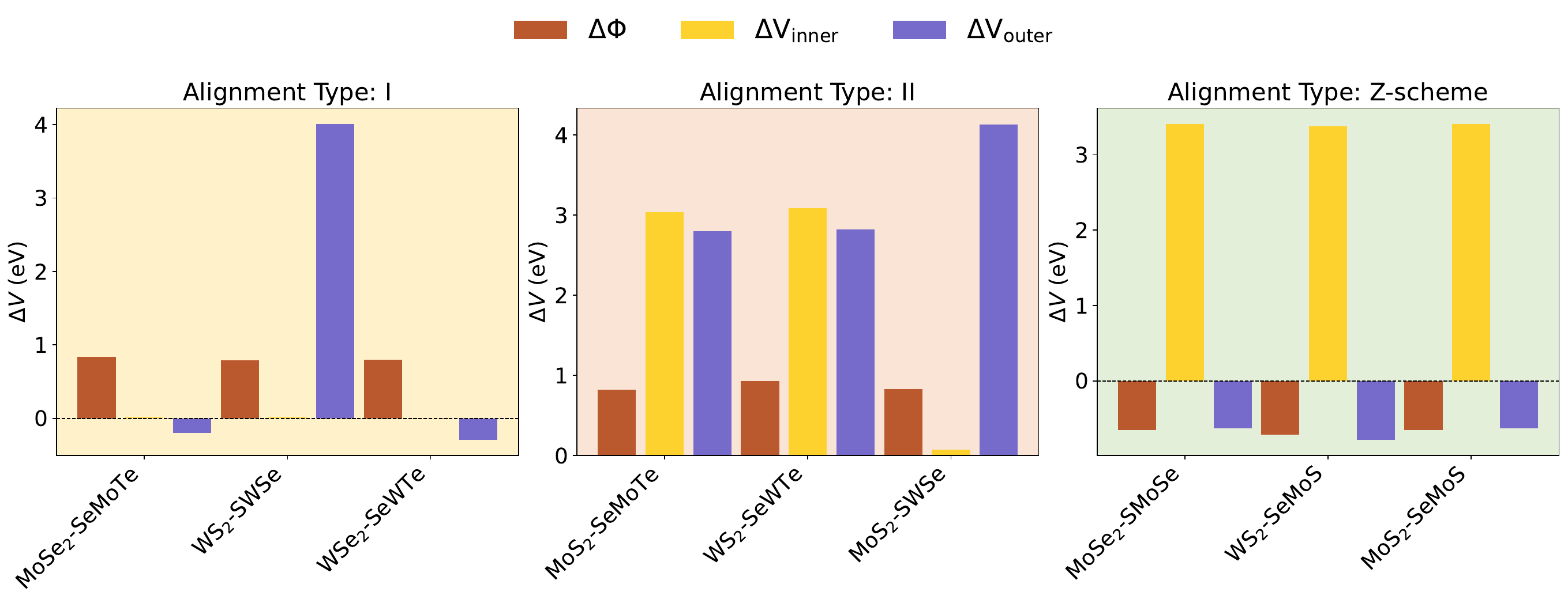}
  \caption{Electrostatic potential differences across interface and vacuum regions in Janus and TMDC sites for Type-I, Type-II, and Z-Scheme alignments.}
  \label{fig:8}
\end{figure*}

As illustrated in Figure~\ref{fig:8} and Figure~S3 (Supporting Information), materials with minimal internal electric fields at the interface tend to exhibit Type-I alignment, especially when the chalcogenide atoms at the interface are identical (i.e., $\Delta V_{\text{inner}} \sim 0$)-with the notable exception of MoS$_2$–SWSe. In contrast, significant internal electric fields, driven by disparities in electronegativity between the contacting chalcogenide atoms, are strongly associated with Type-II and Z-scheme alignments. This means that a strong electric field at the Janus/TMDC interface can separate the CBM and VBM between the Janus and TMDC layers. Recent studies have highlighted that transitions between Type-I, Type-II, and Z-scheme alignments can be induced using external methods, such as applying strain, introducing external electric fields, or modifying interfacial composition via chemical functionalization \cite{rahimi2023interfacial, Yuan2021-ab}. For example, strain engineering has been shown to shift band alignments in ZrIBr/MgClBr and MoSTe/WSTe heterobilayers from Type-I to Type-II or Z-scheme, thereby enhancing their suitability for photovoltaic and catalytic applications \cite{Naik2024-yz}. 

Our findings emphasize the convenience of achieving these transitions intrinsically through electronegativity differences and atomic configurations at the interface. Notably, Type-II alignments generally align the internal electric field direction with the synthetic electric field of the heterostructure (i.e., $\Delta V_{\text{inner}} > 0$ and $\Delta V_{\text{outer}} > 0$), whereas Z-scheme alignments exhibit opposing directions (i.e., $\Delta V_{\text{inner}} > 0$ and $\Delta V_{\text{outer}} < 0$), enhancing charge separation and optimizing performance in photocatalysis (except MoSe$_2$-TeMoSe and WSe$_2$-TeWSe). When Janus and TMDC layers feature different metal atoms, Type-II or Z-scheme configurations are more likely, as the internal electric field aligns with the synthetic electric field of the outer chalcogenides. In contrast, heterobilayers with identical chalcogenide atoms often result in Type-I alignments, which typically confine photogenerated electrons and holes in the same layer, limiting charge separation and transport efficiency. 

The mechanism behind the electric field tunes the band alignment from Type-I to Type-II or Z-scheme has been reported in previous studies, linking it to electron accumulation at the layer surface region \cite{rahimi2023interfacial, Yuan2021-ab}. Specifically, an increase in electron density occurs when the direction of the interfacial electric field shifts from pointing toward the contact region to pointing away from it. This shift leads to the stronger downward bending of the band levels in the surface region of heterobilayers with the electric field directed away from the contact area. Consequently, the band character of the heterobilayers transitions from Type-I to Type-II or Z-scheme as the direction of the interfacial electric field changes.

In photocatalytic applications, the objective is to leverage Z-scheme configurations, which mitigate the drawbacks of narrow intrinsic band gaps in monolayer TMDCs or Janus TMDCs while retaining strong light absorption and leading to the spatial separation of photogenerated electrons and holes. This spatial separation optimizes photocatalytic performance \cite{Shabbir2024-bc, Wang2024-tt}. Type-I heterostructures typically replicate the narrow band gaps of their constituent layers, and Type-II configurations further narrow the overall band gap, reducing their effectiveness in photocatalysis. Given the intrinsic narrow band gaps of monolayer semiconductors like TMDCs, the water-splitting potential of Z-scheme heterostructures becomes crucial. This underscores the necessity of selecting suitable configurations during material screening to achieve optimal alignment for water-splitting photocatalysis. A strategic approach to predicting alignment types involves analyzing the electronegativity of the outermost chalcogenides to determine the direction of the overall electric field. When this field aligns with the internal electric field between layers, a Type-II alignment is likely; otherwise, the structure tends toward a Z-scheme configuration.

These insights offer a robust framework for designing Janus/TMDC heterobilayers with precisely controlled synthetic and internal electric fields, enabling the development of optimized materials. By strategically arranging atomic positions to manipulate electric fields, these heterobilayers can be tailored for specific applications in photocatalysis, light harvesting, and advanced optoelectronic devices \cite{Yuan2021-ab, rahimi2023interfacial}.

\subsection{Solar to Hydrogen Conversion Efficiency (STH)}
After screening all possible materials for their ability to initialize water splitting, we identified 9 materials (see Table~\ref{tab:eff}) with an effective band gap greater than 1.23 eV, indicating that these materials can potentially straddle the whole redox potential of H$_2$O. The redox potential of water depends on pH and can be described as follows \cite{peng2024tunable,ju2019tunable}:
\begin{equation}
    E_{\text{H}^+/\text{H}_2} = -4.44 + 0.059 \times \text{pH},
\end{equation}
and
\begin{equation}
E_{\text{O}_2/\text{H}_2\text{O}} = -5.67 + 0.059 \times \text{pH}.
\end{equation}

Determining the appropriate pH for water splitting is crucial for optimizing the performance of photocatalysts. Therefore, we analyzed the band alignment of the selected materials for the redox potentials at $\text{pH} =0$ and $7$, as illustrated in Figure~S4 (Supporting Information). We find that some materials require solution $\text{pH} < -2$, indicating highly acidic conditions, which limits their practical application \cite{ahmad2016photocatalytic,janani2020recent}. Therefore, these materials are excluded from the STH efficiency calculations. Only materials capable of splitting water at $\text{pH} =0$ and $7$ are considered for the STH efficiency evaluation, with the relevant material information provided in Table~2. It is noted that the VBM of WSe$_2$-SWSe is slightly higher than the OER potential. Therefore, at $\text{pH} =7$, this material exhibits only the hydrogen evolution half-reaction during photocatalysis. Interestingly, all the materials identified as suitable for photocatalytic water splitting exhibit a Z-scheme alignment. From our perspective, this suggests that a large percentage of viable photocatalysts in this group of Janus/TMDC heterobilayers adopt the Z-scheme configuration. This can be attributed to their effective band alignment, which enables direct electron transfer between semiconductors with complementary band structures, making them highly suitable for photocatalytic applications.

\begin{table}[t]
\centering
\caption{STH conversion efficiencies, including overpotentials $(\chi_{\text{H}}$ and $\chi_{\text{O}})$, absorption $(\eta_{\text{abs}})$ and carrier utilization efficiencies $(\eta_{\text{cu}})$ for selected materials.}
\label{tab:eff}
\begin{tabular}{l l l l l l}      
\hline
\textbf{Material} & \textbf{$\chi_{\text{H}}$} & \textbf{$\chi_{\text{O}}$} & \textbf{$\eta_{\text{abs}}$} & \textbf{$\eta_{\text{cu}}$} & \textbf{$\eta_{\text{STH}}$} \\
\hline
WSe$_2$-SWSe 
& 0.10 & 0.01 & 96.63\% & 22.75\% & 15.35\% \\
WSe$_2$-TeWSe 
& 0.12 & 0.25 & 87.08\% & 23.23\% & 15.36\% \\
WS$_2$-SMoSe 
& 0.09 & 0.39 & 88.54\% & 26.39\% & 16.62\% \\
WS$_2$-SeMoS & 0.27 & 0.73 & 60.68\% & 28.81\% & 14.56\% \\
MoS$_2$-SMoSe& \multicolumn{5}{c}{\multirow{5}{*}{Require low pH (pH $<$ 0)}} \\
MoS$_2$-SeMoS&  &  &  &  &  \\
WS$_2$-SWSe&  &  &  &  &  \\
WS$_2$-SeWS &  &  &  &  &  \\
MoS$_2$-SeWS&  &  &  &  &  \\
\hline
\end{tabular}
\end{table}

\begin{table*}[t]
\centering
\caption{Material properties of selected Janus-TMDC heterobilayers, including effective masses ($m^*$), deformation potentials ($E_d)$, calculated carrier mobilities ($\mu_\text{LA}$), calculated mobilities contributed by LO1 and LO2 modes ($\mu_\text{LO1}$ and $\mu_\text{LO2}$) and combined LA and LO modes ($\mu_\text{total}$).}
\label{tab:LA}
\resizebox{\textwidth}{!}{%
\begin{tabular}{cccccccccc} 
\hline
\textbf{Sample} & \textbf{Direction} & \textbf{Carrier} & \textbf{$m*/m_{0}$} & \textbf{$E_{d}$}& \textbf{$C_\text{2D}$} & \textbf{$\mu_\text{LA}$} & \textbf{$\mu_{\text{LO1}}$} & \textbf{$\mu_{\text{LO2}}$} & \textbf{$\mu_\text{total}$}\\ 
& & & & \textbf{(eV)} & \textbf{(N/m)} & \textbf{(cm$^2$/V$\cdot$s)}  
& \textbf{(cm$^2$/V$\cdot$s)} & \textbf{(cm$^2$/V$\cdot$s)} & \textbf{(cm$^2$/V$\cdot$s)}\\
\hline
\multirow{4}{*}{WSe$_2$-SWSe} & \multirow{2}{*}{zigzag}& h & 0.401 & 2.26 & \multirow{2}{*}{218.78} & 5085.70 & 980.78 & 5669.29 & 718.07 \\
 &  & e & 0.225 & 2.70 &  & 11120.74 & 1719.25 & 9937.93 & 1295.01 \\
 & \multirow{2}{*}{armchair}& h & 0.499 & 2.18 & \multirow{2}{*}{201.17} & 4038.86 & 980.78 & 5669.29 & 692.72 \\
 &  & e & 0.290 & 2.64 &  & 8298.47 & 1719.25 & 9937.93 & 1245.67 \\
\hline
\multirow{4}{*}{WSe$_2$-TeWSe} & \multirow{2}{*}{zigzag}& h & 0.611 & 2.52 & \multirow{2}{*}{212.98} & 1735.10 & 946.32 & 246.97 & 175.99 \\
 &  & e & 0.257 & 2.16 &  & 13274.89 & 2237.96 & 584.06 & 447.56 \\
 & \multirow{2}{*}{armchair}& h & 0.743 & 2.46 & \multirow{2}{*}{199.02} & 1399.13 & 946.32 & 246.97 & 171.81 \\
 &  & e & 0.316 & 2.26 &  & 9215.51 & 2237.96 & 584.06 & 441.01 \\
\hline
\multirow{4}{*}{WS$_2$-SMoSe} & \multirow{2}{*}{zigzag}& h & 0.252 & 4.70 & \multirow{2}{*}{245.10} & 3499.11 & 109174.17 & 2137.68 & 1311.06 \\
 &  & e & 0.343 & 4.70 &  & 1697.92 & 72065.71 & 1411.08 & 762.48 \\
 & \multirow{2}{*}{armchair}& h & 0.285 & 4.78 & \multirow{2}{*}{229.13} & 2796.32 & 109174.17 & 2137.68 & 1198.22 \\
 &  & e & 0.480 & 4.34 &  & 1330.22 & 72065.71 & 1411.08 & 678.28 \\
\hline
\end{tabular}
}
\end{table*}

The STH efficiency $\eta_{\text{STH}}$ of the selected materials were calculated using the following expression~\cite{thanh2023janus, Wang2022-xs}:
\begin{equation}
    \eta_{\mathrm{STH}}=  \eta_{\mathrm{abs}} \times \eta_{\mathrm{cu}} \times \frac{\left.\int_0^{\infty} P(\hbar \omega) \mathrm{d}(\hbar \omega)\right)}{\int_0^{\infty} P(\hbar \omega) \mathrm{d}(\hbar \omega)+\Delta\Phi \int_{E_{\mathrm{g}}}^{\infty} \frac{P(\hbar \omega)}{\hbar \omega} \mathrm{d}(\hbar \omega)},
\end{equation}
where $\eta_{\text{abs}}$ is the light absorption efficiency, $\eta_{\text{cu}}$ is the carrier utilization efficiency, $E_g $ is the band gap energy, $\Delta \Phi$ is the electric potential difference, and $P(\hbar \omega)$ represents the AM1.5G solar flux at photon energy $\hbar \omega$. The light absorption efficiency ($\eta_{\text{abs}}$) is given by:
\begin{equation}
\eta_{\mathrm{abs}}=\frac{\int_{E_{\mathrm{g}}}^{\infty} P(\hbar \omega) \mathrm{d}(\hbar \omega)}{\int_0^{\infty} P(\hbar \omega) \mathrm{d}(\hbar \omega)},
\label{equ:8}
\end{equation}
, the carrier utilization $\eta_{\text{cu}}$ is determined by:
\begin{equation}
\eta_{\mathrm{cu}}=\frac{1}{2} \times \frac{\Delta E_\text{spliting} \int_{E_{\min }}^{\infty} \frac{P(\hbar \omega)}{\hbar \omega} \mathrm{d}(\hbar \omega)}{\int_{E_{\mathrm{g}}}^{\infty} P(\hbar \omega) \mathrm{d}(\hbar \omega)},
\end{equation}
where $\Delta E_\text{spliting}$ = 1.23 eV for water splitting and $E_{\text{min}}$ is defined as
\begin{equation}
E_{\text{min}} =
\begin{cases} 
0 & \text{for}~~ \chi\left(\mathrm{H}_2\right) \geq 0.2, \chi\left(\mathrm{O}_2\right) \geq 0.6 \\
E_{\mathrm{g}} + 0.2 - \chi\left(\mathrm{H}_2\right) & \text{for}~~ \chi\left(\mathrm{H}_2\right) < 0.2, \chi\left(\mathrm{O}_2\right) \geq 0.6 \\
E_{\mathrm{g}} + 0.6 - \chi\left(\mathrm{O}_2\right) & \text{for}~~ \chi\left(\mathrm{H}_2\right) \geq 0.2, \chi\left(\mathrm{O}_2\right) < 0.6 \\
E_{\mathrm{g}} + 0.8 - \chi\left(\mathrm{H}_2\right) - \chi\left(\mathrm{O}_2\right) & \text{for}~~ \chi\left(\mathrm{H}_2\right) < 0.2, \chi\left(\mathrm{O}_2\right) < 0.6
\end{cases},
\end{equation}
where $\chi(\text{H}_2)$ and $\chi(\text{O}_2)$ are the overpotentials for the hydrogen and oxygen evolution reactions, respectively. It is noted that $\chi(\text{H}_2)$ is the potential difference between the CBM and the redox potential of $\text{H}^+/\text{H}_2$, while $\chi(\text{O}_2)$ is the potential difference between the VBM and the redox potential of $\text{O}_2/\text{H}_2\text{O}$, as shown in Figure~\ref{fig:2}. The factor 1/2 in the carrier utilization efficiency formula implies that only half of the photoinduced carriers (electrons and holes) are effectively utilized for the HER and OER in the Z-scheme system. This is because the Z-scheme mechanism involves the recombination of the electrons from the CBM of one material with the holes in the VBM of another material at the interface \cite{Wang2022-xs}. In particular, WS$_2$-SMoSe achieves the highest $\eta_{\text{STH}}$ of 16.62\%. This demonstrates a significant improvement compared to previously reported Z-scheme vdW materials. For example, MoSe$_2$/Ti$_2$CO exhibits an STH efficiency of 12\% \cite{Fu2021-yn}, while MoSe$_2$/SnSe$_2$ and WSe$_2$/SnSe$_2$ achieve efficiencies of 10.5\% and 9.3\%, respectively \cite{Fan2019-wm}. Additionally, CrS$_3$/GeSe achieves 10.60\% \cite{Wan2023-pd}, and ZrS$_2$/SeMoS reaches 6.3\% \cite{Wang2022-lu}. Furthermore, the predicted STH efficiency surpasses the critical threshold of 10\% required for commercial applications, highlighting its potential as a promising photocatalyst for hydrogen production. 

Although the efficiency of the Z-scheme is only half that of Type II or Type I alignments, its importance becomes evident when the intrinsic band gap of the materials narrows. The Z-scheme enables these materials to function as photocatalysts by aligning their VBM and CBM to straddle both the HER and OER potentials, as reported in \citeauthor{Wang2022-xs}'s works. The promising performance of WS$_2$-SMoSe and other materials in Table~\ref{tab:eff} can be attributed to its optimal electronic structure and favorable band alignment, which facilitate efficient light absorption and charge carrier utilization. The light absorption efficiency $\eta_{\text{abs}}$ is particularly high for materials with smaller band gaps, as shown in Eq.~\ref{equ:8}. WSe$_2$-SWSe with a band gap of 0.604 eV exhibits an outstanding $\eta_{\text{abs}}$ of 96.63\%, which allows it to absorb a broader spectrum of solar radiation compared to traditional photocatalysts with a standard band gap of 1.23 eV. 

It is important to note that  Eq.~\ref{equ:8} assumes the absorbance as a step function zero for photon energies below the band gap and unity for energies above it. This simplification may lead to underestimating $\eta_{\text{abs}}$, particularly at lower photon energies. Therefore, while the calculated $\eta_{\text{abs}}$ provides valuable insights, it may be more accurate for materials with high photon energy absorption. 


\begin{figure*}[t]
  \centering
  \includegraphics[width=1\linewidth]{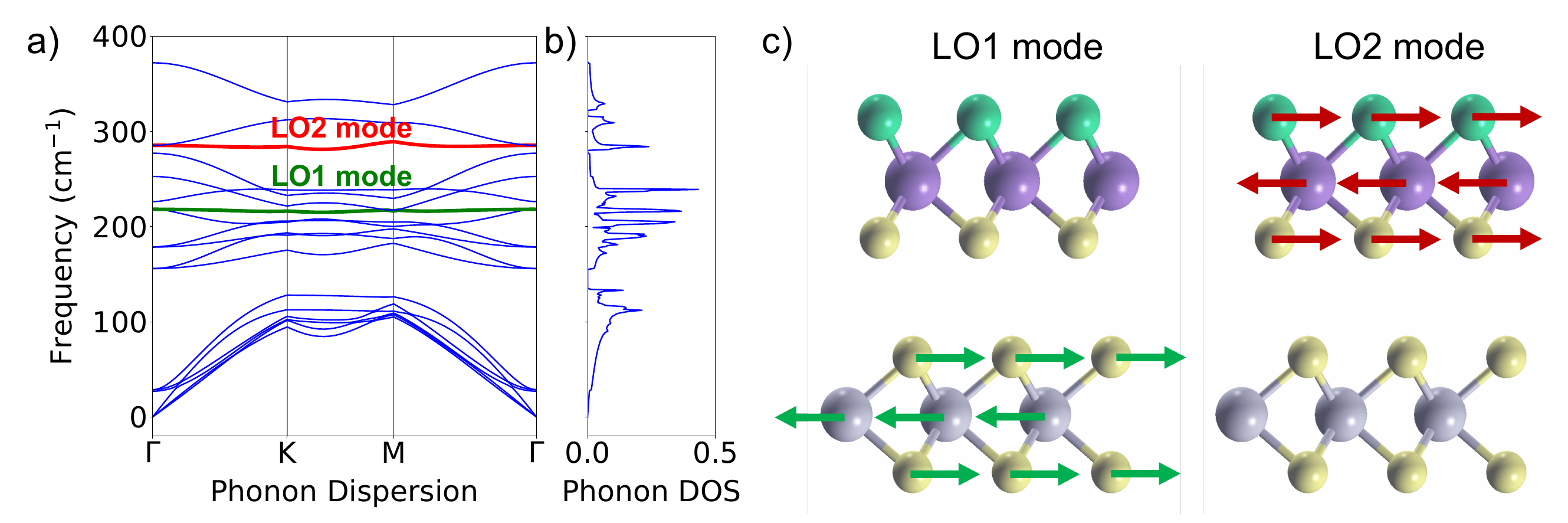}
  \caption{Phonon properties of W$\text{Se}_2$-SWSe (a) Phonon dispersion, (b) Phonon density of state (DOS), and (c) the vibrations of the LO1 and LO2 phonon modes at $\Gamma$ point.}
  \label{fig:phonon}
\end{figure*}

\subsection{Precision and efficiency approach in transport calculation}
The relationship between carrier mobility and photocatalytic efficiency is essential in determining the performance of photocatalysts. After the electron-hole pairs are separated, the free photoexcited carriers need to migrate to the chemical active sites for water splitting. Higher mobility enables these charge carriers to travel faster and further, reducing the probability of recombination before they can reach the reactive surface sites \cite{lv2017two, liu2020two, qiao2018germanium, ma2024carrier}. Mobility is primarily limited by phonon scattering, which impedes the movement of charge carriers. Traditional approaches to calculating mobility, such as deformation potential theory (DPT), focus on the scattering caused by longitudinal acoustic (LA) phonons \cite{yuan2023first, ma2024carrier, thanh2023janus}. However, Cheng and Liu~\cite{cheng2018limits} have demonstrated that this method often underestimates the influence of longitudinal optical (LO) phonon scattering, particularly in Janus materials with strong polar effects.

The carrier mobility $\mu_\text{LA}$ due to the LA phonon scattering within the DPT theory is given as follows~\cite{ma2024carrier, thanh2023janus}:
\begin{equation}
    \mu_\text{LA}=\frac{e \hbar^3 C_{2 \mathrm{D}}}{k_{\mathrm{B}}T m^*_{x(y)} m_{\mathrm{d}} E_{\mathrm{d}}^2},
\end{equation}
where the effective masses along the zigzag ($x$) and armchair ($y$) directions are represented by $m_x^*$ and $m_y^*$, respectively, and the average effective mass given by $m_d = \sqrt{m_x^* m_y^*}$. A $\sqrt{3} \times 1$ supercell (see Figure~S5 in Supporting Information) is used to compute $m_x^*$ and $m_y^*$. The in-plane Young's modulus, $C_{2\mathrm{D}}$, is defined as~\cite{ma2024carrier, thanh2023janus}:
\begin{equation}
C_{2\mathrm{D}} = \frac{\partial^2 E_{\text{total}}}{\partial \varepsilon^2}/ S_0,
\end{equation}
where $E_{\text{total}}$ is the total energy after applying a uniaxial strain (with strain $\varepsilon = \Delta l / l_0$), and $S_0$ is the equilibrium area of the system (see Figure~S6, Supporting Information). The deformation potential constant, $E_d$, is given by
\begin{equation}
E_d = \frac{\Delta E}{\varepsilon},
\end{equation}
where $\Delta E$ is the shift in the energy level of either the CBM or the VBM relative to the vacuum level (see Figure~S7 in Supporting Information). The calculated parameters at room temperature ($T = 300$ K) are presented in Table~\ref{tab:LA}.

Recent advancements in understanding the intrinsic mobility of 2D TMDCs have emphasized the critical role of LO phonon scattering, in addition to the traditionally considered LA phonon scattering \cite{cheng2018limits}. To accurately capture these effects, many previous studies have utilized full electron-phonon coupling (EPC) calculations based on density functional perturbation theory (DFPT) combined with Wannier interpolation \cite{Vu2023-al, Sohier2018-eu, Zhang2023-cb, Zhao2018-pq}. This approach enables precise computation of EPC matrix elements over dense k- and q-point grids, effectively overcoming the limitations of DPT, which assumes isotropic electron-phonon coupling and predominantly accounts for LA phonon scattering. However, while the EPC method offers a more accurate and comprehensive depiction of phonon-limited mobility, it is computationally intensive and demands substantial resources for high-precision simulations \cite{Zhang2022-ud}.

A more practical approach is to separately calculate the contributions of LO and LA phonon modes and subsequently determine the total mobility using Matthiessen’s rule \cite{cheng2018limits}. This can be achieved by employing the DPT formula for LA phonon scattering and the formula for Fr\"{o}hlich interaction in polar 2D materials~\cite{sohier2016two, cheng2018limits}. Including LO phonon scattering is particularly vital for polar materials, as its strength is closely linked to the Born effective charge, which quantifies the degree of electric polarization change caused by atomic vibrations. A larger Born effective charge results in stronger carrier scattering and, consequently, reduced mobility. This insight highlights the necessity of advancing beyond conventional DPT to accurately evaluate intrinsic mobility, especially for Janus-TMDC heterobilayers, where LO phonon scattering significantly influences carrier transport properties. The total scattering rate near the band edges can be written as

\begin{equation}
\begin{split}
\frac{1}{\tau_\text{LO}}&=\frac{ e^4 \hbar n_{\hbar \omega}}{M_M M_{X1} M_{X2} S_0 d_\text{eff}^2 (\varepsilon\varepsilon_0)^2(\hbar \omega)^2}\times \\
    &~~~~~\left[ \left|Z_{M B} \right| \sqrt{M_{X1}M_{X2}} + \left| Z_{X 1}\right| \sqrt{M_{M}M_{X2}} + \left|Z_{X2} \right|\sqrt{M_{M}M_{X1}}\right]^2,
\end{split}
\label{tauLO}
\end{equation}
where $\tau_\mathrm{LO}$ represents the relaxation time, quantifies the rate at which charge carriers lose momentum due to phonon scattering, \( e \) is the elementary charge, \( n_{\hbar \omega} \) is the Bose-Einstein distribution at the phonon frequency at the gamma point, \( A \) is the equilibrium area, \( d_\text{eff} \) is the effective thickness, \( \varepsilon \) is the in-plane optical dielectric constant, and \( \varepsilon_0 \) is the dielectric constant. Additionally, \( M_M \), \( M_{X1} \), and \( M_{X2} \) are the atomic masses of the metal and the two chalcogens in the structure, while \( Z_{MB} \), \( Z_{X1} \), and \( Z_{X2} \) are the Born effective charges of the metal and the two chalcogen atoms(see Table~S2 in Supporting Information for reference). Thus, the mobility contributed by the LO phonon mode is given by
\begin{equation}
\mu_\text{LO}={|e| \tau_\text{LO}}/{m^*}. 
\label{muy13}
\end{equation}

In Janus-TMDC heterobilayers, two distinct LO phonon modes, LO1 and LO2, correspond to the vibrations of the TMDC and Janus layers, respectively, as shown in Figure~\ref{fig:phonon}. The calculated mobilities for $\mu_\text{LO1}$ and $\mu_\text{LO2}$ are provided in Table~\ref{tab:LA}. Then the total mobility $\mu_\text{total}$ is then determined using Matthiessen’s rule as follows \cite{cheng2018limits}:
\begin{equation}
\frac{1}{\mu_\text{total}} = \frac{1}{\mu_\text{LO1}} + \frac{1}{\mu_\text{LO2}} + \frac{1}{\mu_\text{LA}}.
\end{equation} 
The calculated $\mu_\text{LO1}$, $\mu_\text{LO2}$, and $\mu_\text{total}$ are listed in Table~~\ref{tab:LA}. Our results reveal that LO phonon scattering significantly limits mobility in Janus-TMDC heterobilayers, which leads to lower carrier mobility compared to those predicted by DPT alone (see Table~~\ref{tab:LA}). Importantly, our approach to calculating the mobility of TMDC materials and their Janus heterobilayers proves to be not only more precise but also computationally less expensive than methods relying solely on DPT or EPC. This balance between accuracy and efficiency underscores the robustness of our methodology in capturing the intricate phonon-electron interactions in these systems.

\subsection{Enhanced mobility in Janus/TMDC heterobilayers}
In the approach using the Fr\"{o}hlich model to calculate the mobility of polar materials, Cheng and Liu~\cite{cheng2018limits} compared the mobilities of monolayer MoS\(_2\), MoSe\(_2\), WS\(_2\), and WSe\(_2\) calculated via Fr\"{o}hlich scattering and the EPC method~\cite{cheng2018limits}. Their findings revealed that the mobilities in these materials are primarily influenced by the LO and LA phonon modes. However, compared to the Fr\"{o}hlich mobility of Janus/TMDC heterobilayers listed in Table~\ref{tab:LA}, our heterobilayers exhibit significantly higher mobilities than those of monolayer TMDCs~\cite{cheng2018limits}. This enhancement in mobility can be attributed to three main factors: (1) dielectric screening from the other layers, (2) remote phonon scattering from distant polar-optical phonons, and (3) changes in the Fr\"{o}hlich scattering strength due to polar-optical phonon coupling \cite{Zhang2024-jx}.

Additionally, we find that in TMDC/Janus heterobilayers, the internal electric field modifies the Born effective charges, reducing their magnitude and thereby increasing the Fr\"{o}hlich mobility. For instance, the Born effective charges of W atoms in monolayer WSe\(_2\) and WS\(_2\) are $-0.65$ and $-1.29$, respectively~\cite{cheng2018limits}, whereas those in WS\(_2\)-SMoSe, WSe\(_2\)-TeWSe, and WSe\(_2\)-SWSe heterobilayers are $-0.097$, $-0.52$, and $-0.57$, respectively. Another contributing factor is the effective thickness, \( d_\text{eff} \). As shown in Eq.~\ref{tauLO} and \ref{muy13}, the increasing \( d_\text{eff} \) enhances the Fr\"{o}hlich mobility. Consequently, the heterobilayers achieve higher mobilities compared to their monolayer counterparts. Furthermore, very recent corrections to the polar-optical phonon interaction in the Fr\"{o}hlich model demonstrated that the Fr\"{o}hlich mobility of InSe in an InSe/BN vdW heterostructure was enhanced by $2.5$ times compared to that of monolayers of InSe \cite{Zhang2024-jx}.

This significant mobility enhancement in heterobilayers compared to monolayers suggests that the Z-scheme alignment of our heterobilayers offers a new insight advantage: not only does it provide a favorable alignment for photocatalytic water splitting by reducing the electron-hole recombination rate, but it also significantly amplifies mobility, making them superior to monolayers.

\begin{figure}[t]
  \centering
  \includegraphics[width=0.5\linewidth]{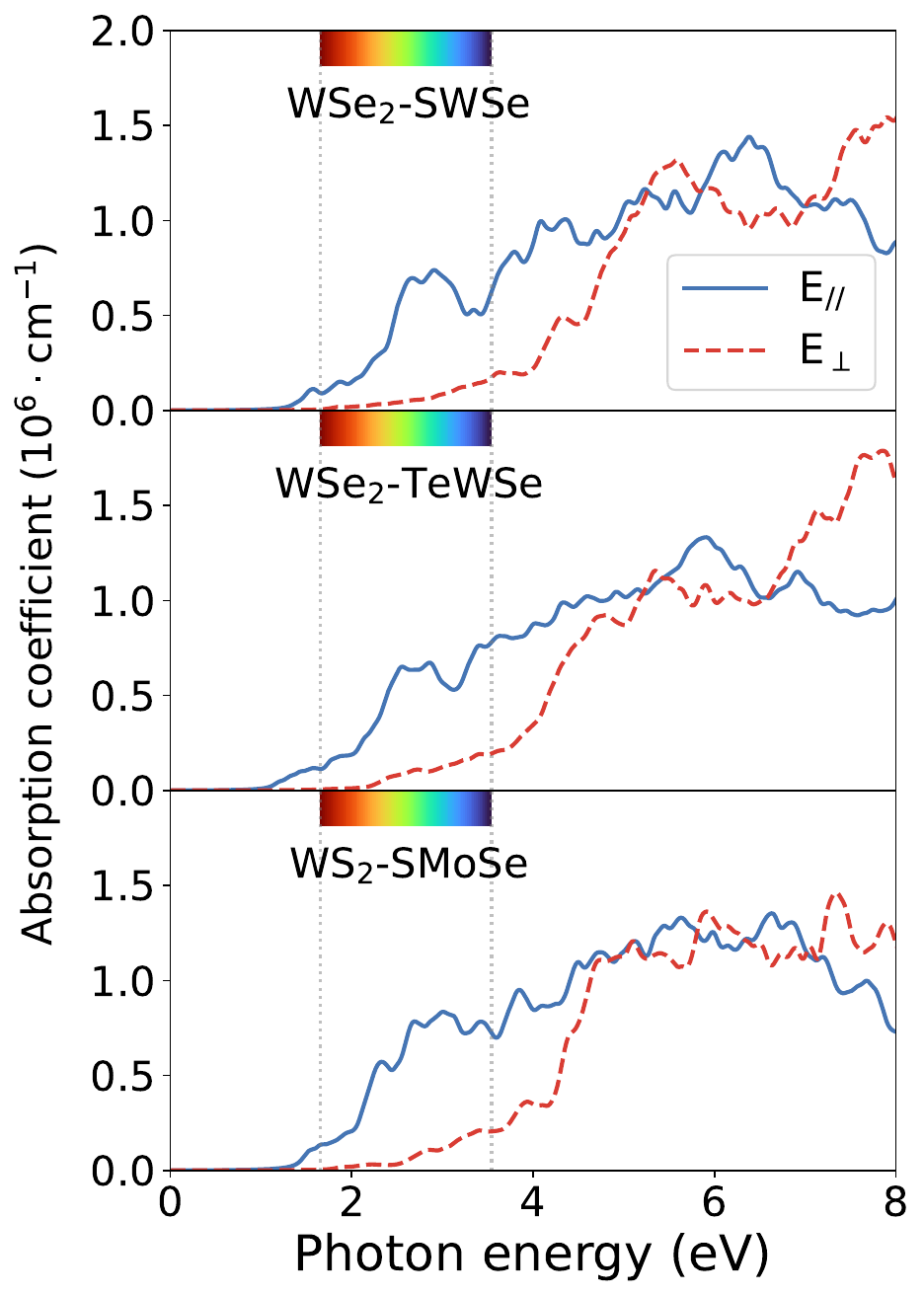}
  \caption{Optical absorption spectra along in-plane ($E_\parallel$) and out-of-plane ($E_\perp$) of WSe$_2$-SWSe, WSe$_2$-TeWSe, and WS$_2$-SMoSe. The dashed line denotes the range of the visible light.}
  \label{fig:optics}
\end{figure}

\subsection{Optical properties}
The optical property is another factor in determining the effectiveness of semiconductor materials as optoelectronic devices. Additionally, photocatalysts with strong light absorption capabilities will perform more efficiently.

The optical properties of $\mathrm{WSe_2}$-SWSe, $\mathrm{WSe_2}$-TeWSe, and $\mathrm{WS_2}$-SMoSe are calculated by using the independent particle approximation. The absorption coefficient $\alpha(\omega)$ is obtained from the calculated real $\varepsilon_1$ and imaginary $\varepsilon_2$ parts of the dielectric function as follows~\cite{hung2022quantum}:
\begin{equation}
    \alpha(\omega) =\frac{\sqrt{2} \omega}{c}\left\{\sqrt{\varepsilon_1^2(\omega)+\varepsilon_2^2(\omega)}-\varepsilon_1(\omega)\right\}^{1 / 2}.
\end{equation}
Here, the values of $\varepsilon_1$ and $\varepsilon_2$ are scaled by the factor $h/d_0$ where $h$ represents the unit cell thickness and $d_0$ denotes the effective material thickness \cite{yuan2023first}. The effective thickness, $d_0$, is calculated using the formula: $d_0 = d_{\text{outer}} + r_{\text{atom1}} + r_{\text{atom2}}$, where $d_{\text{outer}}$ is the distance between the two outermost atoms of the heterostructure, as illustrated in Figure~\ref{fig:1}. The terms $r_{\text{atom1}}$ and $r_{\text{atom2}}$ correspond to the vdW radii of the two outermost atoms \cite{alvarez2013cartography}.

In our study, we observed that the optical absorption spectra of TMDC/Janus heterobilayers exhibit strong anisotropic behavior. The in-plane absorption coefficient $\alpha(\omega)$ ($E_\parallel$) reaches up to $0.84 \times 10^6$ cm$^{-1}$ in the visible light region for $\mathrm{WS_2}$-SMoSe and WSe$_2$-TeWSe, and $0.77 \times 10^6$ cm$^{-1}$ for WSe$_2$-SWSe. These values significantly surpass those of monolayer Janus materials such as $\gamma$-GeSSe ($0.44 \times 10^6$ cm$^{-1}$) \cite{thanh2023janus} and PdSSe ($0.24 \times 10^6$ cm$^{-1}$) \cite{guo2023strain}, as well as monolayers of MoSSe, MoSeTe, MoSTe, WSSe, WSeTe, and WSTe~\cite{xia2018universality}. This substantial enhancement and broadening of absorption in TMDC/Janus heterobilayers can be attributed to strong light–matter coupling \cite{Lin2024-kb, Munkhbat2019-ah}, which is more pronounced in multilayered structures. 

Notably, the in-plane optical absorption ($E_\parallel$) differs considerably from the out-of-plane absorption ($E_\perp$), emphasizing the anisotropic optical properties of these heterobilayers. The combination of superior STH efficiency and strong light absorption across all materials underscores their potential for optoelectronic and water-splitting applications. This highlights the role of the Z-scheme configuration for narrow band gap semiconductors, which creates favorable conditions for the water-splitting process by extending the intrinsic band gap while retaining the efficient light-harvesting properties of small band gap materials.

\subsection{Hydrogen Evolution Reaction}
\subsubsection{Gibbs free energy steps for HER}
The research indicates that the selected photocatalysts exhibit characteristics of direct Z-scheme systems, featuring appropriate band edge potentials and high efficiency. However, these properties alone do not ensure the spontaneous occurrence of the water-splitting reaction under light irradiation. To further evaluate the hydrogen evolution reaction (HER) capabilities, the Gibbs free energy change ($\Delta G_\text{H}$) for hydrogen adsorption is considered a crucial factor in assessing the HER capabilities of photocatalysts. To accurately calculate the adsorption energy and minimize interactions between adjacent atoms, a 3 $\times$ 3 $\times$ 1 supercell was constructed for each material, ensuring that the results reflect the behavior of isolated hydrogen atoms. The Gibbs free energy change was determined using the equations: 
\begin{equation}
    \Delta G_\text{H} = \Delta E_\text{H} + 0.24  \text{ eV},
\end{equation}
and
\begin{equation}
\Delta E_\text{H} = E_\text{Janus-TMDC + H} - E_\text{Janus-TMDC} - \frac{1}{2}E_\mathrm{H_2},
\end{equation}
where $E_{\text{Janus-TMDC + H}}$ is the total energy of the heterostructure of Janus monolayer and TMDC material with an adsorbed hydrogen atom, $E_{\text{Janus-TMDC}}$ is the energy of the bare vdW layers, and $\frac{1}{2}E_\mathrm{H_2}$ represents the energy of a hydrogen molecule in its standard state.
Building upon this, the potential provided by photogenerated electrons for HER ($U_e$) can be calculated using the equation $U_e = V_{\text{CB}} - E_{\text{H}^+/\text{H}_2} $ \cite{yuan2023first, ju2019tunable}, where $U_e$ corresponds to the electron potential associated with the CBM, potentially matching the electron affinity $\chi(\text{H}_2)$ under illuminated conditions. In a dark environment, where no photogenerated electrons are available, $U_e$ becomes zero, indicating the absence of a driving force for HER. Based on the calculations in the previous section, the \( U_e \) values for $\text{WS}_2$-$\text{SMoSe}$, $\text{WSe}_2$-$\text{TeWSe}$, and $\text{WS}_2$-$\text{SMoSe}$ under the illumination were found to be 0.10, 0.12, and 0.09 eV, respectively, at pH = 7. At pH = 0, the corresponding values were 0.52 eV, 0.53 eV, and 0.51 eV, respectively.

\begin{figure*}[t]
  \centering
  \includegraphics[width=1\linewidth]{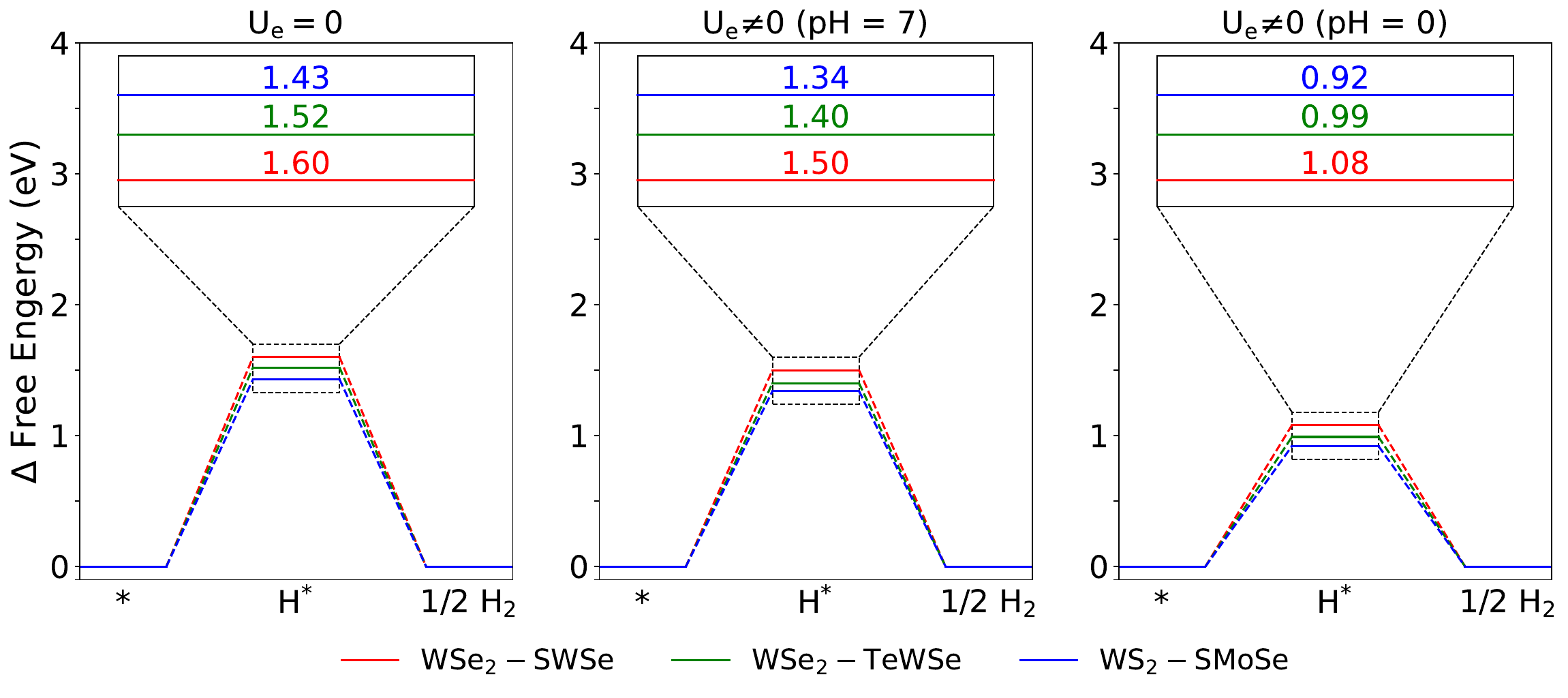}
  \caption{Calculated Gibbs free energy steps for HER under various illumination conditions.}
  \label{fig:h2}
\end{figure*}

We further evaluated the HER activity at the reduction sites of \( \text{WSe}_2 \)-\( \text{SWSe} \), \( \text{WSe}_2 \)-\( \text{TeWSe} \), and again \( \text{WS}_2 \)-\( \text{SMoSe} \), with the corresponding \( \Delta G_\text{H} \) values presented in Figure~\ref{fig:h2}. The calculated free energy changes without an external potential are 1.60, 1.52, and 1.43~eV for these materials, respectively. These high positive values of \( \Delta G_\text{H} \) indicate that spontaneous hydrogen adsorption does not occur easily, implying limited catalytic efficiency under standard conditions. Nevertheless, this behavior is similar to other well-known materials such as \( \text{MoS}_2 \), \( \text{MoSe}_2 \), \( \text{WSe}_2 \), and the Janus \( \text{MoSSe} \) materials \cite{xiao2021high, tang2016mechanism,tsai2014active}. On the other hand, calculated free energy changes with an external potential are the \( \Delta G_\text{H} \) values to 1.50, 1.40, and 1.34~eV for \( \text{WSe}_2 \)-\( \text{SWSe} \), \( \text{WSe}_2 \)-\( \text{TeWSe} \), and \( \text{WS}_2 \)-\( \text{SMoSe} \), respectively, at pH = 7. At pH = 0, the corresponding values were 1.08 eV, 0.99 eV, and 0.92 eV, respectively. These reduction values suggest that suitable irradiation can enhance the HER performance by decreasing the energy barrier for hydrogen adsorption. Although the \( \Delta G_\text{H} \) values remain positive, indicating that the hydrogen evolution reaction is still non-spontaneous, the decreased values under light exposure or applied potential could potentially improve the catalytic performance by making hydrogen releasing more favorable.

\begin{figure}[t]
    \centering
    \includegraphics[width=0.7\linewidth]{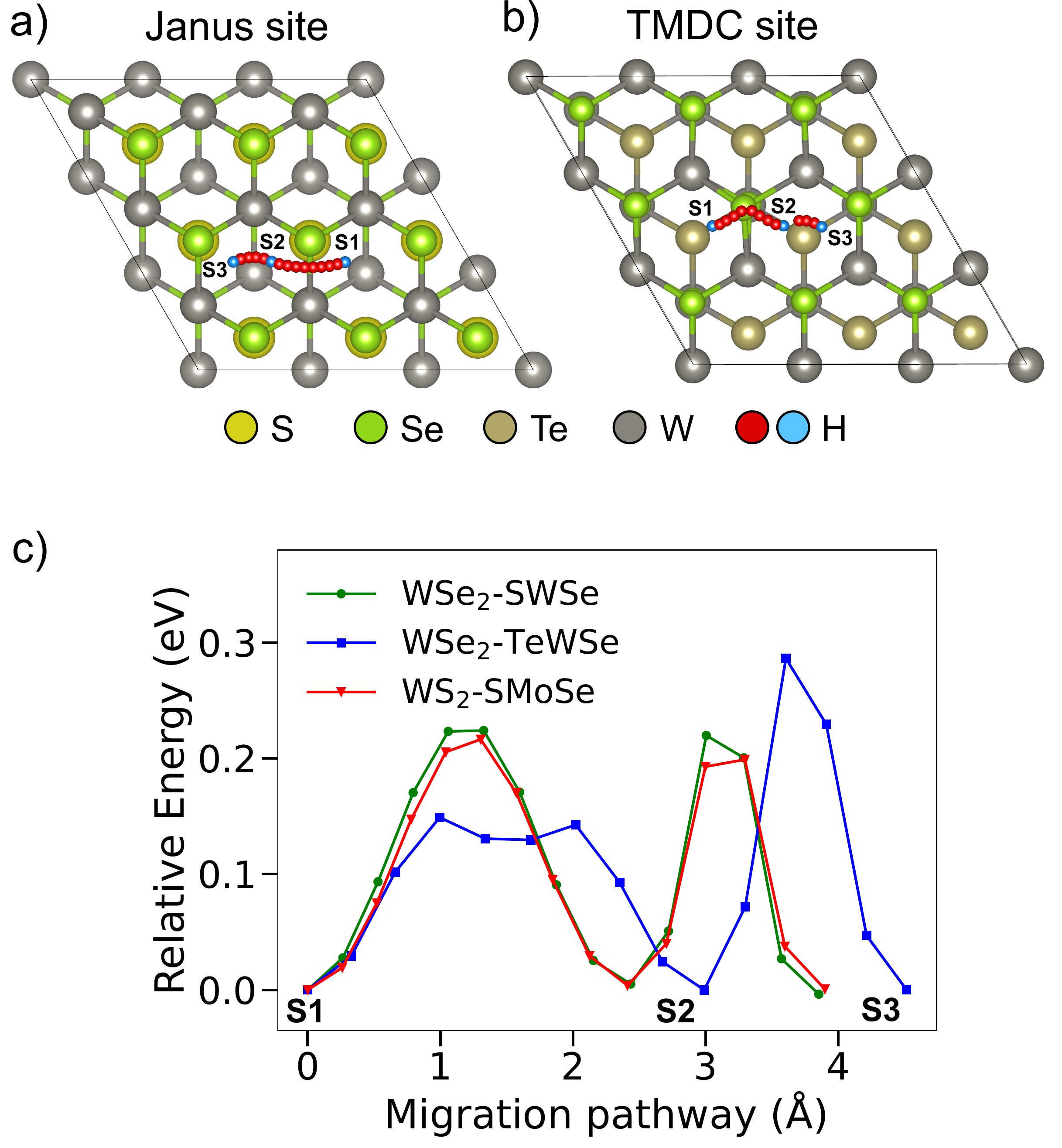}
    \caption{Migration pathways and energy profiles of a Hydrogen atom on the reduction site: (a) Janus site (WSe$_2$-SWSe and WS$_2$-SMoSe) and (b) TMDC site (WSe$_2$-TeWSe). S1 $\rightarrow$ S2 indicates migration between neighboring hex-atomic-rings, while S2 $\rightarrow$ S3 represents migration within a hex-atomic-ring.}
    \label{fig:6}
\end{figure}

\subsubsection{Diffusion of H atom on the HER sites}
To investigate H atom diffusion on the HER-active surfaces, we calculate the optimized migration pathways using a minimal energy approach \cite{lei2021first}. The pathways and the corresponding energy profiles are shown in Figure~\ref{fig:6}, where S1 $\rightarrow$ S2 represents H atom migration between adjacent hex-atomic-rings, and S2 $\rightarrow$ S3 denotes migration within a single hex-atomic-ring. The migration pathways are optimized by the nudged elastic band (NEB) method \cite{lei2021first} with a convergence criterion of 0.05 eV/\r{A} for NEB calculations with 13 intermediate images. For WS$_2$-SMoSe, H migration on the SMoSe surface showed a lower energy barrier for inter-ring migration (0.199 eV) than intra-ring migration (0.216 eV). In the case of WSe$_2$-TeWSe, the WSe$_2$ surface exhibited easier migration between hex-atomic-rings (0.149 eV) than within a ring (0.286 eV). For WS$_2$-SMoSe, the SMoSe surface favored intra-ring migration (0.200 eV) over inter-ring movement (0.224 eV), with this structure exhibiting a lower energy barrier compared to the other two. 
The preference of the H atom for the Janus site over the TMDC site can be attributed to the smaller electronegativity difference between the H atom and the S atom (on the Janus site) compared to the difference between the H atom and the Se atom (on the TMDC site). This distinction highlights the potential of designing Janus/TMDC heterobilayers with minimal interaction between the H atom and chalcogenide atoms on the reduction site, thereby reducing the interaction between the H atom and the material layers. Such design could lead to more efficient catalytic performance.

Compared to previously reported values \cite{lei2021first}, the migration energy barriers of these structures are lower than those of monolayer MoSSe, which were observed at 0.32 eV for inter-ring and 0.08 eV for intra-ring migration on the Se site. This reduction suggests that the WS$_2$-SMoSe, WSe$_2$-TeWSe, and WS$_2$-SMoSe surfaces provide more favorable diffusion pathways. Furthermore, thanks to the Z-scheme heterostructure alignment, the electric field was modified, which contributed to reducing the interaction between the H atom and the photocatalyst system, as observed in the 3 calculated samples \cite{Bhagat2023-zj}. This modification plays a role in enhancing the efficiency of the HER.

\section{Conclusions}
The findings in this study underscore the potential of Janus-TMDC heterobilayers as highly efficient photocatalysts for STH conversion. By leveraging intrinsic electric fields generated through Janus asymmetry, we achieve enhanced charge separation and minimized electron-hole recombination in Z-scheme configurations such as WSe$_2$-SWSe, WSe$_2$-TeWSe, and WS$_2$-SMoSe. These Z-scheme structures not only enable spatially separated sites for HER and OER but also provide favorable band alignment for water splitting, with WS$_2$-SMoSe achieving the highest STH efficiency of 16.62\%. Our framework highlights the critical role of electronegativity differences and interfacial atomic configurations in determining band alignment types. Materials exhibiting significant internal electric fields and aligned synthetic and internal electric field directions favor Type-II configurations, while opposing field directions lead to Z-scheme alignments. This predictive framework aids in the rational design of heterobilayers with tailored electronic properties, advancing their application in photocatalysis and beyond. In exploring the relationship between carrier mobility and photocatalytic capabilities, we use the Fr\"{o}hlich interaction model to account for LO phonon scattering, revealing that mobility in Janus-TMDC heterobilayers is significantly affected by their intrinsic polarization. Additionally, HER analysis shows that external potential effectively lowers reaction barriers, enhancing HER efficiency under illumination. Hydrogen diffusion also indicates relatively low energy barriers for H atom migration on HER-active surfaces, suggesting efficient diffusion on the WSe$_2$-SWSe, WSe$_2$-TeWSe, and WS$_2$-SMoSe surfaces. Overall, this work not only establishes a comprehensive framework for evaluating and predicting the performance of Janus-TMDC heterobilayers but also paves the way for innovative design strategies. By integrating insights from band alignment, intrinsic electric fields, and carrier mobility, we present a holistic approach to optimizing these heterobilayers for clean energy applications. The findings provide a robust foundation for advancing next-generation photocatalytic systems, offering transformative potential for hydrogen production, light harvesting, and other energy-related technologies.

\begin{acknowledgement}
This work was financially supported by Vietnam National University Ho Chi Minh City (NCM2024-50-01).
\end{acknowledgement}

\begin{suppinfo}
   \textbf{ Supporting Information:} Table of structural parameters and band gap; table of Born charges, Phonon frequency, and dielectric properties; figure of HSE06-calculated band structure; figure of electrostatic potential along vacuum axis ($z$); figure of electrostatic potential differences across interface for different alignment types; figure of band alignments for water splitting at pH 0, 2, and 7; figure of PBE band structure for the effective mass calculation; figure of total energy-strain relationship in zigzag and armchair directions; figure of CBM and VBM energies under uniaxial strain; figure of Phonon dispersion and DOS.

\end{suppinfo}
\providecommand{\latin}[1]{#1}
\makeatletter
\providecommand{\doi}
  {\begingroup\let\do\@makeother\dospecials
  \catcode`\{=1 \catcode`\}=2 \doi@aux}
\providecommand{\doi@aux}[1]{\endgroup\texttt{#1}}
\makeatother
\providecommand*\mcitethebibliography{\thebibliography}
\csname @ifundefined\endcsname{endmcitethebibliography}  {\let\endmcitethebibliography\endthebibliography}{}

\end{document}